\documentstyle[e-e-ijmpa,epsfig]{article}

\newcommand{\lesssim}{\mathop{}_{\textstyle \sim}^{\textstyle <}}
\newcommand{\gtrsim}{\mathop{}_{\textstyle \sim}^{\textstyle >}}  
\newcommand{\emem}{e^-e^-}
\newcommand{\epem}{e^+e^-}
\newcommand{\ifb}{~{\rm fb}^{-1}}
\newcommand{\mev}{~{\rm MeV}}
\newcommand{\gev}{~{\rm GeV}}
\newcommand{\tev}{~{\rm TeV}}

\newcommand{\selectron}{\tilde{e}}
\renewcommand{\thefootnote}{\fnsymbol{footnote}} 

\begin{document}

\normalsize\textlineskip

\noindent
\begin{minipage}[t]{\textwidth}
\begin{flushright}
hep-ph/9803319 \hfill
LBNL--41254 \\
March 1998 \hfill
UCB--PTH--98/04 \\
\end{flushright}
\end{minipage}
\vspace*{.3in}

\title{\LARGE \bf
Supersymmetry at Linear Colliders:\\
The Importance of Being \boldmath$\emem$}

\author{\large Jonathan L. Feng
\footnote{This work was supported in part by the Director, 
Office of Energy Research, Office of High Energy and Nuclear Physics,
Division of High Energy Physics of the DOE under Contracts
DE--AC03--76SF00098 and by the NSF under grant PHY--95--14797.}  
}

\address{\normalsize
\vspace*{0.2in}
Theoretical Physics Group, 
Lawrence Berkeley National Laboratory \\
and Department of Physics, University of California\\
Berkeley, CA 94720 \, USA}

\maketitle

\vspace*{0.4in}

\abstracts{\normalsize
Advantages of the $\emem$ option at linear colliders for
the study of supersymmetry are highlighted.  The fermion number
violating process $\emem \to \selectron^- \selectron^-$ provides
unique opportunities for studies of slepton masses and flavor mixings.
In particular, slepton mass measurements at the 100 MeV level through
threshold scans of scalar pair production may be possible. Such
measurements are over an order of magnitude better than those possible
in $\epem$ mode, require far less integrated luminosity, and may lead
to precise, model-independent measurements of $\tan\beta$.
Implications for studying gauginos and the importance of accurate beam
polarimetry are also discussed.
{\normalsize
\vspace*{0.25in}
\begin{center}
To appear in\\
\vspace*{0.07in}
Proceedings of the 2nd International Workshop on\\
Electron-Electron Interactions at TeV Energies\\
University of California, Santa Cruz, 22--24 September 1997
\end{center}}
}

\setcounter{footnote}{0}
\renewcommand{\thefootnote}{\alph{footnote}}

\vspace*{1pt}\textlineskip	

\newpage

\section{Introduction}

The possible role of supersymmetry (SUSY) in stabilizing the
electroweak scale is cause for optimism in the search for SUSY at
current and planned colliders.  If SUSY is discovered, detailed
studies of superpartner properties will likely become a long-term
focus of high energy physics and the primary goal of future colliders.

In recent years, our appreciation for the variety of possible
superpartner mass spectra, flavor structures, and SUSY breaking
mechanisms has grown dramatically.  At future colliders, it will
therefore be important to seek {\em model-independent} measurements of
all possible superpartner properties.  Such studies will yield
constraints on SUSY parameters that ultimately could shed light on a
variety of mysteries, including the physics at or near the Planck
scale.

What contributions might an $\emem$ collider make toward this goal?
The replacement of a beam of positrons with electrons is
straightforward at linear colliders, and the option of colliding
electrons in the $\emem$ mode is therefore, for the most part, a
simple extension for any linear collider program.  However, when
considering the physics promise of the $\emem$ mode, it is, of course,
important first to recall the potential of the more conventional
hadron or $\epem$ colliders.  In particular, a direct comparison can
be made to the $\epem$ mode of linear colliders, where luminosities of
$50\ifb/{\rm yr}$, center of mass energies of up to 1.5 TeV, and
highly polarizable $e^-$ beams have been shown to be powerful tools
for model-independent studies of SUSY particles.  At $\epem$
colliders, superparticles may be discovered essentially up to the
kinematic limit, and their couplings may be measured at the percent
level to determine if they are, in fact, supersymmetric partners of
standard model particles.\cite{HN,FMPT,NFT,CFP,KRS,NPY} Detailed
studies of the chargino and neutralino sectors,\cite{JLC,FMPT,BMT}
sleptons,\cite{BV,JLC,NFT,BMT,Bartl} and
squarks\cite{squark,BMT,Bartl} find that the masses of most of these
particles may be measured to a few percent, and mixings, such as
gaugino-Higgsino mixing\cite{JLC,FMPT} and left-right scalar
mixing,\cite{NFT,Bartl} may also be determined.

What, then, can an $\emem$ collider add?  At first sight, there appear
to be only disadvantages.  In $\emem$ mode, pair production of almost
all superpartners is forbidden by total lepton number and charge
conservation:

\begin{equation}
e^-e^- \not\to \chi^-\chi^-, \chi^0\chi^0, \tilde{q}\tilde{q}^*,
\tilde{\nu} \tilde{\nu} \ .
\label{otherpairs}
\end{equation}
It is therefore clear that the general purpose potential of $\epem$
colliders cannot be matched by $\emem$ colliders.  In fact, the only
possible superpartner pair production is the fermion number violating
process\cite{KL}

\begin{equation}
e^-e^- \to \selectron^-\selectron^- \ ,
\label{selectron}
\end{equation}
which is allowed through the $t$-channel Majorana gaugino exchange of
Fig.~\ref{fig:feynman}.  The advantages of the $\emem$ mode over the
$\epem$ mode for SUSY studies are almost certainly confined to those
derived from this reaction.

\begin{figure}[t]
\centerline{\epsfig{file=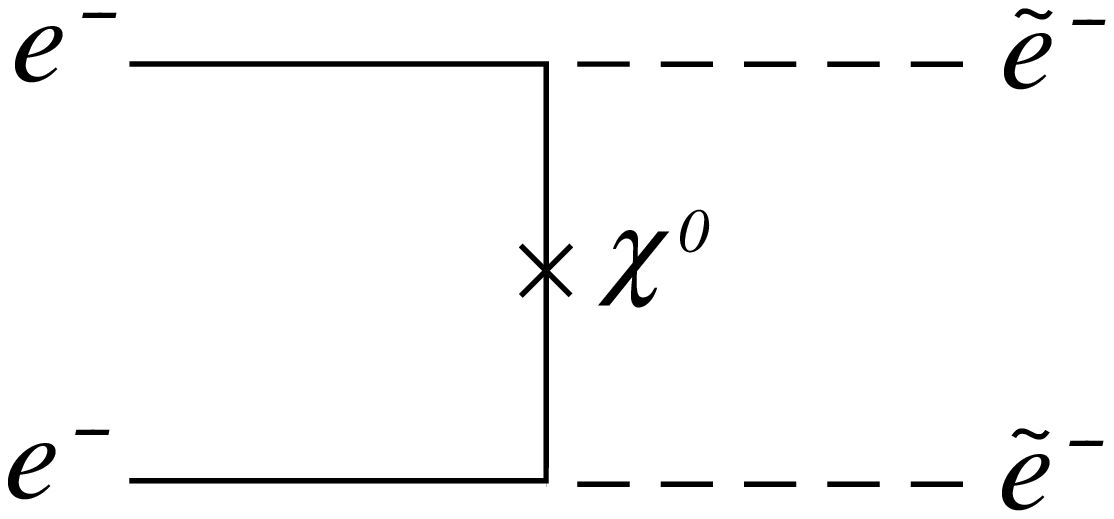,width=0.7\textwidth}}
\vspace*{0.1in}
\fcaption{The contribution to $e^- e^- \to \selectron^- \selectron^-$
from $t$-channel Majorana neutralino exchange.}
\label{fig:feynman}
\end{figure}

The process $e^- e^- \to \selectron^- \selectron^-$, however, is
particularly well-suited to precision studies.  First, backgrounds may
be highly suppressed.  Second, selectrons are typically expected to be
among the lighter superparticles, and they are therefore likely to be
kinematically accessible. As we will see, the cross sections for
$e^-e^- \to \selectron^- \selectron^-$ are then typically large, and
so statistical errors are small.  Third, the properties of selectrons
are largely determined by quantum numbers, and so selectron production
and decay have strong dependences on only a few SUSY parameters.
Theoretical systematic errors arising from unknown SUSY parameters are
therefore also typically small.

In fact, the only selectron properties not determined by quantum
numbers are their masses and flavor mixings, and, as we will see in
the following two sections, $\emem$ colliders provide unparalleled
potential for detailed studies of both of these properties.  Note,
however, that the simple characteristics of selectrons also make them
ideal for probing other sectors.  A few comments on implications for
gaugino mass measurements will be given below.\cite{PT} In
addition, high precision measurements of selectron couplings may be
used to constrain very massive sparticle sectors through the
super-oblique corrections introduced in Ref.~\citenum{CFP} --- this
possibility is described in the contribution of H.--C. Cheng to these
proceedings.\cite{Cheng}

\section{Slepton Masses}

Let us consider first the case of $\selectron_R$ pair production in
the absence of flavor mixing.  At an $\epem$ collider, this takes
place through $s$-channel $\gamma$ and $Z$ processes and $t$-channel
neutralino exchange. Assuming that the selectron decays directly to a
stable neutralino $\chi$, the signal is $\epem \to \epem \chi \chi$,
where the neutralinos go undetected.  The dominant backgrounds are
$W^+W^-$, which can be nearly eliminated by right-polarizing the $e^-$
beam, and $e^{\pm} \nu_e W^{\mp}$ and $\gamma\gamma \to W^+ W^-$, which
cannot.

As the reaction requires a right-handed electron and a left-handed
positron, the initial state has spin 1, leading to the well-known
$\beta^3$ behavior of scalar pair production at threshold.
Measurements of scalar masses through threshold scans are therefore
impossibly poor, and one must resort to kinematic endpoints.  For
example,\cite{BV,JLC,BMT} the upper and lower endpoints of the energy
distributions of the final state $e^+$ and $e^-$ are determined by
$m_{\selectron_R}$ and $m_{\chi}$, and by measuring these endpoints,
$m_{\selectron_R}$ may be constrained to a few GeV with an integrated
luminosity of 20 to 50 $\ifb$.\footnote{In such analyses, the
information contained in the fact that electrons and positrons come
paired in events is lost.  Using kinematic variables that are
sensitive to this correlation,\cite{squark} slepton mass measurements
may be improved, sometimes very significantly.\cite{Wagner,Lykken}
These improved analyses do not reduce the required integrated
luminosities, however, and measurements much below the GeV level still
appear to be rather challenging.}

In the $\emem$ mode, selectron pair production takes place only
through $t$-channel neutralino exchange.  The signal is $\emem \to
\emem \chi\chi$.  However, among the potential backgrounds, $W^-W^-$
is forbidden by total lepton number conservation, $\gamma\gamma \to
W^+W^-$ does not produce like-sign electrons, $e^- e^- Z$ may be
eliminated by kinematic cuts,\cite{COR} and the remaining backgrounds
$e^- \nu_e W^-$ and $\nu_e \nu_e W^- W^-$ may be completely
eliminated, in principle, by right-polarizing {\em both} beams.

In addition, the initial state $e^-_R e^-_R$ required for
$\selectron^-_R \selectron^-_R$ production has spin 0, and the
threshold cross section therefore has the $\beta$ behavior more
commonly associated with fermion pair production.  The $\sqrt{s}$
dependence of the cross section is shown in
Fig.~\ref{fig:crosssection} for $m_{\selectron_R} = 200\gev$, where,
for simplicity, we have assumed gaugino-like neutralinos, and the
effects of initial state radiation, beamstrahlung, and selectron width
have been neglected.  For comparison, the $\epem$ cross section is
also plotted; it is barely visible near threshold.

\begin{figure}[t]
\centerline{\epsfig{file=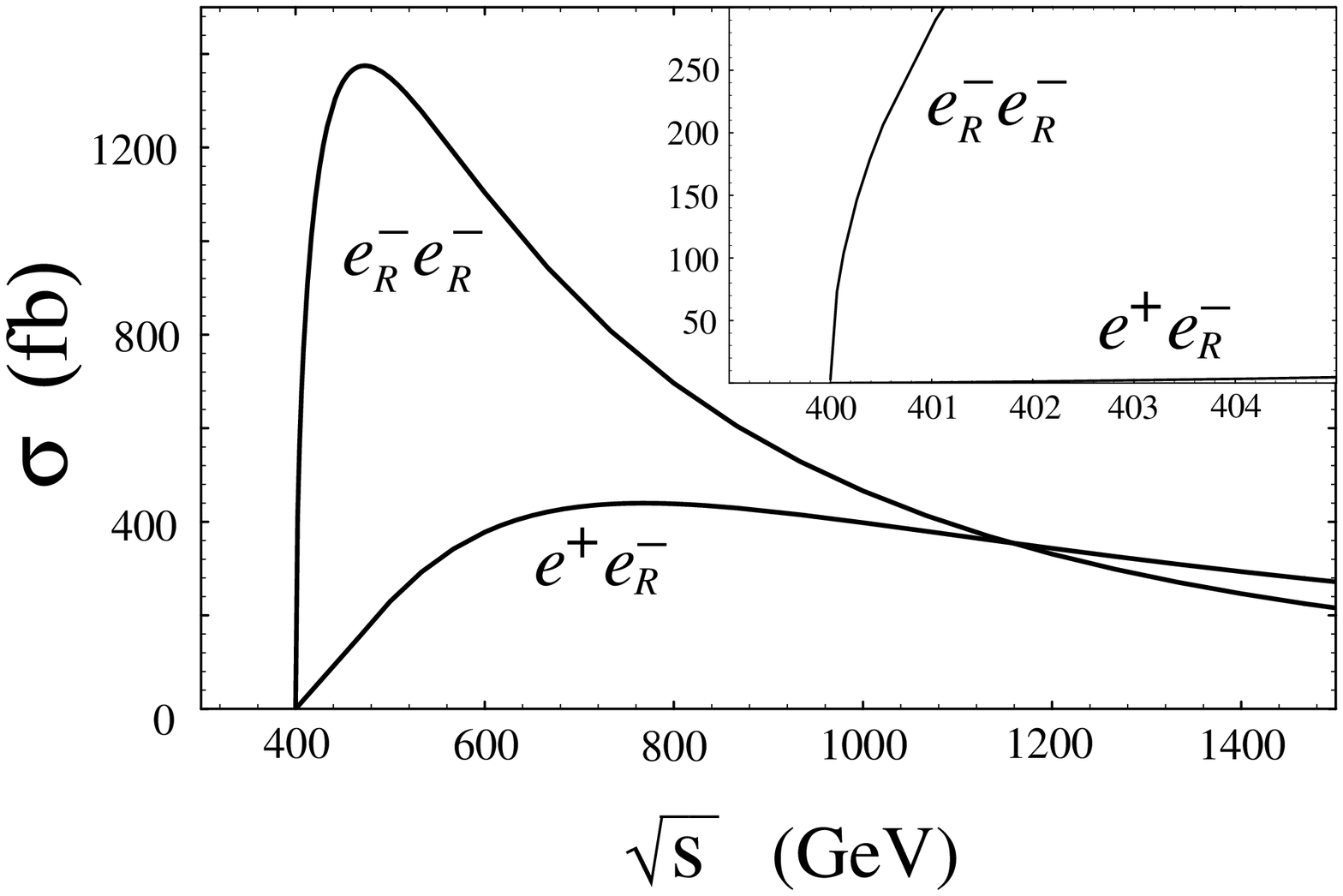,width=0.88\textwidth}}
\vspace*{0.1in}
\fcaption{Cross sections $\sigma(e^-_R e^-_R \to \selectron^-_R
\selectron^-_R)$ and $\sigma(e^+e^-_R \to \selectron^+_R
\selectron^-_R)$ for $m_{\selectron_R} = 200 \gev$ and $m_{\tilde{B}}
= 100\gev$. The inset is a magnified view for $\sqrt{s}$ near
threshold.  Effects of initial state radiation, beamstrahlung, and the
selectron width are not included.}
\label{fig:crosssection}
\end{figure}

As the $\emem$ cross section rises sharply at threshold, let us now
consider what precision might be expected from a threshold mass
measurement.  The $1\sigma$ statistical error on the mass from a
measurement of the cross section is

\begin{equation}
\Delta m = \Delta \sigma \left( \frac{\partial \sigma}{\partial m}
\right)^{-1} \ ,
\end{equation}
where $\Delta \sigma = \sqrt{\sigma/L}$, and $L$ is the total
integrated luminosity.  At $\sqrt{s} = 2m_{\selectron_R} + \,
0.5\gev$, where the cross section is $\sigma = 200$ fb, an integrated
luminosity of $L = 1\ifb$ gives a cross section measurement of $\Delta
\sigma = 14$ fb, and the resulting $1\sigma$ statistical uncertainty
on the mass is $\Delta m = 40 \mev$.  This result contrasts sharply
with results from the $\epem$ mode, which, as noted above, are
typically more than an order of magnitude worse.  Note also that the
necessary integrated luminosity can be collected in a matter of weeks,
even given the possible factor of 2 to 3 reduction in luminosity for
the $\emem$ mode relative to the $\epem$ mode.\cite{Spencer}

In the above, effects of background have been neglected.  The dominant
background arises from imperfect beam polarization, and is $e^-\nu_e
W^-$ with cross section $B = 43 \times 2P(1-P) + 400 \times (1-P)^2$
fb.\cite{CC} The beam polarization $P$ is defined here as the fraction
of right-handed electrons in each individual beam: $P = N(e^-_R) /
[N(e^-_L) + N(e^-_R)]$.  Polarizations of $P=90\%$ are already
available, and higher polarizations may be possible for future
colliders.\cite{Clendenin} For $P = 90\% \ (95\%)$, the background is
$B = 12 \ (5)$ fb and is negligible, assuming it is well-understood
and so contributes only to the the uncertainty through statistical
fluctuations.  While the difference between 90\% and 95\% polarization
is not critical for this study, one might worry that the systematic
uncertainty from beam polarization measurement might be significant.
For example, to take an extreme case, if $P = 90\pm 5\%$, the
$e^-\nu_e W^-$ background is constrained only to the range 5 to 20
fb. However, if the projected beam polarization uncertainties of
$\Delta P \sim 1\%$ are achieved,\cite{Woods} the systematic
uncertainty does not significantly degrade these results.

The analysis above is clearly highly idealized, and more concrete
estimates require a number of refinements.\cite{FP} In particular,
effects of the selectron width, initial state radiation, and
beamstrahlung must be included, and other experimental systematic
errors, such as uncertainties in the beam energy, will also be
important at this high level of precision. In addition, theoretical
systematic errors from uncertainties in the masses and gaugino purity
of the neutralinos also enter. Finally, the entire scan must be
optimized once all these effects are included. It is clear, however,
that the $\emem$ mode offers an exceptionally promising method for
measuring selectron masses.

Although the analysis for right-handed selectrons is the most elegant,
other slepton masses may also be measured using the $\emem$ and
$e\gamma$ modes and similar strategies. For example,
$m_{\selectron_L}$ can be measured through $e^-_L e^-_L \to
\selectron^-_L \selectron^-_L$.  In this case, beam polarization may
not be used to remove the dominant backgrounds, but again, if
systematic uncertainties are small, the $\beta$ behavior may be
exploited to obtain a precise measurement.  (Note that $\epem \to
\selectron_R^{\pm} \selectron_L^{\mp}$ also has $\beta$ behavior at
threshold.)  Finally, along similar lines, the cross sections for
chargino pair production $\epem \to \chi^+ \chi^-$ and the $(-,-)$
helicity component of $e^- \gamma \to \tilde{\nu}_e \chi^-$ also rise
as $\beta$ near threshold, and, as noted in Refs.~\citenum{BHK} and
\citenum{BBH}, this behavior may be exploited to determine
$m_{\chi^{\pm}}$ and $m_{\tilde{\nu}_e}$ accurately.\footnote{Note
that this method may be used to measure $m_{\tilde{\nu}_e}$ even if
$\tilde{\nu}_e$ decays invisibly.} \ In this way, all first generation
slepton masses may be measured to high precision.

It is appropriate to ask what use such high accuracy measurements
might be.  One important application is to loop-level SUSY
studies.\cite{HN,NFT,CFP,KRS,NPY,Cheng} Another is to the measurement
of $\tan\beta$, which has important implications for Yukawa couplings,
unification scenarios, and a wide variety of other SUSY measurements.
At tree level, the relation

\begin{equation}
m^2_{\selectron_L} - m^2_{\tilde{\nu}_e} = - M^2_W \cos 2\beta
\label{splitting}
\end{equation}
provides a model-independent measurement of $\tan\beta$.  If these
slepton masses are measured and their mass splitting is highly
constrained, bounds on $\tan\beta$ may be obtained. As an example, in
Fig.~\ref{fig:tanbsoft}, upper and lower bounds are given as a
function of the underlying value of $\tan\beta$ for fixed
$m_{\tilde{\nu}_e} = 200 \gev$ and uncertainties in $m_{\selectron_L}
- m_{\tilde{\nu}_e}$ as indicated.  For moderate and large
$\tan\beta$, $\cos 2\beta \approx -1$, and so constraints from
Eq.~(\ref{splitting}) require high precision measurements of the mass
splitting. We see that if the mass difference is known to, say, 200
MeV, the mass splitting provides a powerful determination of
$\tan\beta$ for $\tan\beta \lesssim 10$.  Note that model-independent
measurements of $\tan\beta$ in the intermediate range $4\lesssim
\tan\beta \lesssim 10$ are extremely difficult; previous suggestions
have been limited to those exploiting processes involving heavy Higgs
scalars.\cite{higgs,FM}

\begin{figure}[t]
\centerline{\epsfig{file=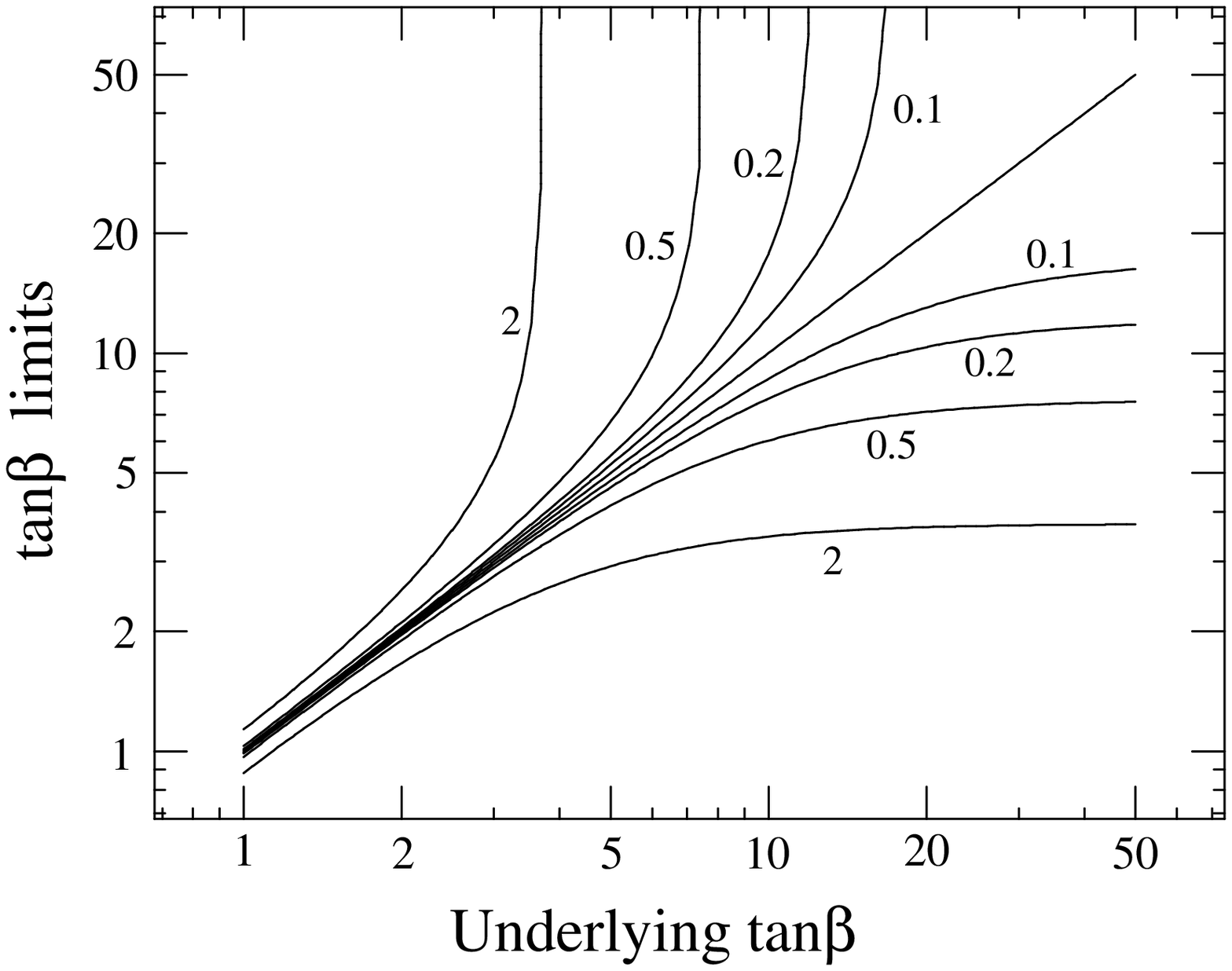,width=0.75\textwidth}}
\vspace*{0.1in}
\fcaption{Contours giving the upper and lower limits on $\tan\beta$ for
a given underlying $\tan\beta$ and experimental uncertainty in mass
difference $\Delta m \equiv m_{\selectron_L} - m_{\tilde{\nu}_e}$ as
indicated (in GeV), for fixed $m_{\tilde{\nu}_e} = 200$ GeV.}
\label{fig:tanbsoft}
\end{figure}

\section{Slepton Flavor Mixings}

In SUSY theories, there are generically many new sources of flavor
violation.  In the standard model, there is no flavor violation at
neutral gauge boson vertices $V^{\mu}\bar{f}f$.  However, this is not
the case for neutral gaugino vertices $\tilde{f}f\tilde{V}$, as the
fermion- and sfermion-diagonalizing matrices need not be identical.
There are therefore 7 new Cabibbo-Kobayashi-Maskawa-like matrices, one
for each fermion species $f=u_L, u_R, d_L, d_R, e_L, e_R, \nu$, all of
which are worthwhile to explore at future colliders.  For simplicity
here, let us consider right-handed lepton flavor violation, and let us
simplify still further to the case of only $\selectron_R -
\tilde{\mu}_R$ mixing, which may be parametrized by a single mixing
angle $\theta_R$.

The mixing of $\selectron_R - \tilde{\mu}_R$ induces decays $\mu\to e
\gamma$ at low energies, and so is already constrained by the rather
stringent bound $B(\mu\to e\gamma ) < 4.9\times 10^{-11}$.\cite{mu}
With the simplifying assumptions above, $\mu\to e\gamma$ receives
contributions from two diagrams, which interfere destructively. Both
are proportional to $(\Delta m^2_R / m^2_R) \sin 2\theta_R$, where
$\Delta m^2_R \equiv m_{\selectron_R}^2 - m_{\tilde{\mu}_R}^2$ and
$m^2_R \equiv (m_{\selectron_R}^2 + m_{\tilde{\mu}_R}^2)/2$, and one
has an additional dependence on the left-right mass mixing parameter
$\hat{t} \equiv (-A+\mu \tan\beta) / m_R$.  Note that the superGIM
suppression factor $\Delta m^2_R / m^2_R$ suppresses the rate for
$\Delta m_R \lesssim m_R$.

The collider signal of lepton flavor violation is $\epem \to e^{\pm}
\mu^{\mp} \chi\chi$ for the $\epem$ mode, or $\emem\to e^- \mu^-
\chi\chi$ for the $\emem$ mode.  In $\epem$ mode, the leading
backgrounds are once again $W^+W^-$, $e\nu_e W$, and $\gamma\gamma \to
W^+W^-$.  The essential virtue of the $\emem$ mode for this study is
the absence of analogous backgrounds if both $e^-$ beams are
right-polarized.

For the $\emem$ case, the flavor-violating collider cross section
takes a form familiar from $B$ physics, and is proportional to $\sin^2
2 \theta \frac{x^2}{1+x^2}$, where $x\equiv \Delta m_R / \Gamma$ and
$\Gamma$ is the slepton decay width.  Note that this cross section is
superGIM suppressed only for $\Delta m_R \lesssim \Gamma$, in contrast
to the $\mu\to e \gamma$ signal.  There is therefore a large range of
mass splittings $\Gamma \lesssim \Delta m_R \lesssim m_R$ where the
low energy signal is suppressed below current bounds, but the collider
signal can be maximally flavor-violating.

In Fig.~\ref{fig:talklfvfig03} we present the reach of an $\emem$
collider in the $(\Delta m^2_R / m^2_R, \sin 2\theta_R)$ plane, where
we demand a 5$\sigma$ excess, and assume $\sqrt{s} = 500 \gev$,
$L=20\ifb$, and 200 GeV right-handed sleptons.  We see that lepton
flavor violation may be probed down to mixing angles $\sim 10^{-2}$,
far below the Cabibbo angle, and for a wide range of mass splittings.
This result is a significant improvement over the $\epem$
case.\cite{ACFH} Note that the discovery of lepton flavor violation
would have major consequences for SUSY models.  For example, the cases
of pure gauge-mediated SUSY and pure minimal supergravity would both
be eliminated, as both assume degenerate sleptons at some scale and
therefore predict the complete absence of lepton flavor violation. For
details, see Refs.~\citenum{ACFH} and \citenum{CPviol}.

\begin{figure}[t]
\centerline{\epsfig{file=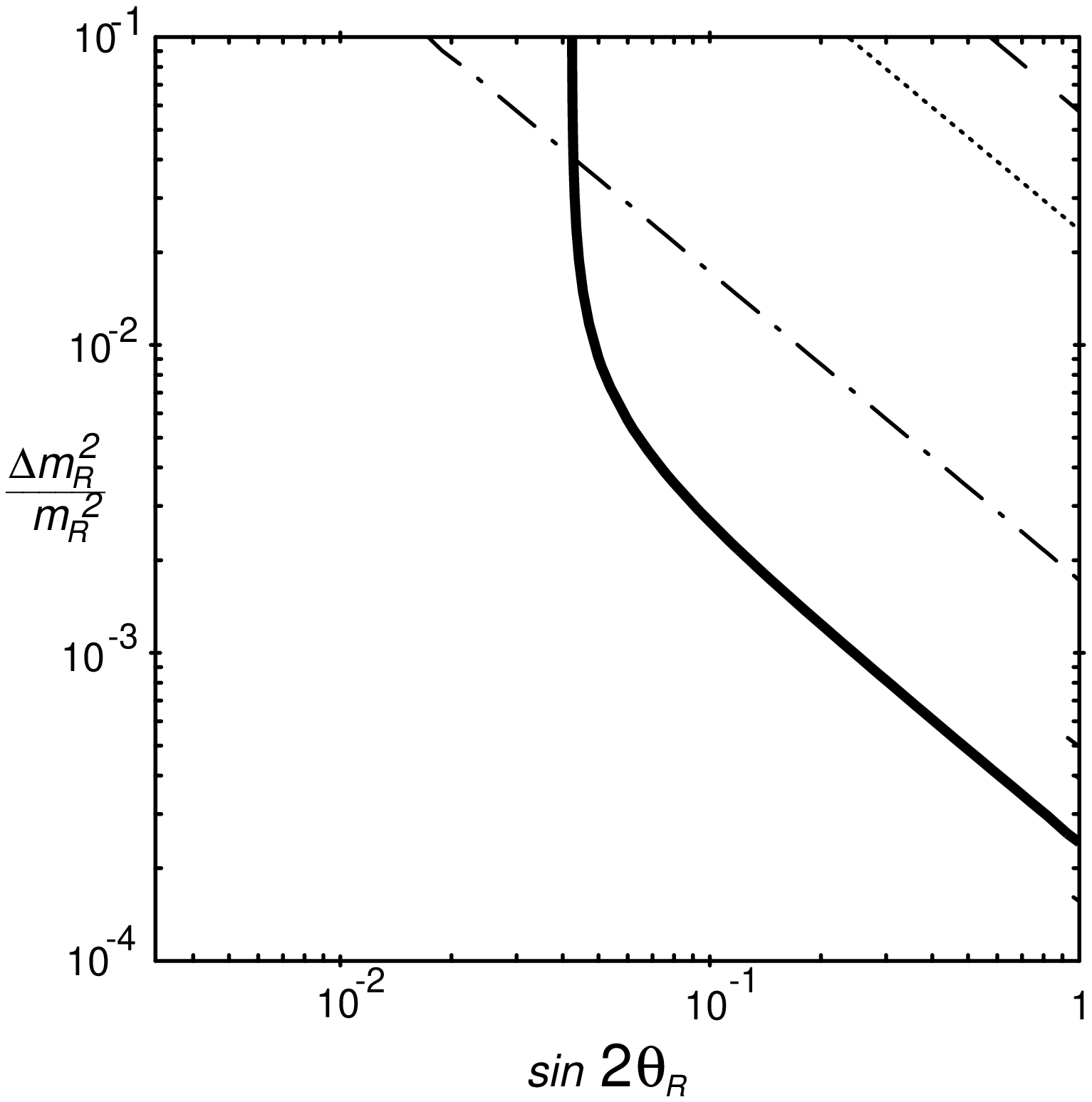,width=0.65\textwidth}}
\vspace*{0.1in} 
\fcaption{The discovery reach (solid) for lepton flavor violation
through the signal $e^-_R e^-_R \to e^- \mu^- \chi \chi$ for 200 GeV
slepton masses, $m_{\chi} = 100 \gev$, $\sqrt{s} = 500\gev$, and an
integrated luminosity $L=20\ifb$. Regions of the plane to the
upper-right are excluded by the current bound $B(\mu\to e\gamma ) <
4.9\times 10^{-11}$ for $\hat{t} = 0$ (dotted), 2 (dashed), and 50
(dot-dashed), where we have assumed $m_{\tilde{l}_L} = 350\gev$.  }
\label{fig:talklfvfig03}
\end{figure}

It is important to note that in presenting these results, we have
assumed a right-handed beam polarization of $P=90\%$, for which the
background is $B=12$ fb and a $5\sigma$ signal is $S=3.9$ fb, and we
have neglected experimental systematic uncertainties in beam
polarization. However, for this study, as we are looking for a rare
signal and a large background has been eliminated through beam
polarization, accurate polarimetry is absolutely crucial.  For
example, as noted above, if the beam polarization is $P=90\pm 5 \%$,
the background is constrained only to the range 5 to 20 fb; the
$5\sigma$ signal is then overwhelmed by polarimetry uncertainties. In
fact, even for $P=90\pm 1\%$, the background ranges from $B = 10$ to
13 fb, which is also significant relative to the statistical
uncertainty. As these SUSY flavor studies may provide important
insights into not only the mixings of superpartners, but also the
observed patterns of standard model fermion masses and mixings, they
are an important example of studies for which beam polarimetry plays
an essential role.

\section{Gaugino Mass Measurements}

As noted in the introduction, the simplicity of selectrons allows one
to use selectrons to probe other sectors.  It is possible, for
example, to exploit the spin structure of the amplitude of
Fig.~\ref{fig:feynman} to study the gaugino sector.  In particular,
because this amplitude includes a $t$-channel neutralino mass
insertion,

\begin{equation}
\sigma (e^-_R e^-_R \to \selectron^-_R \selectron^-_R ) \sim \left|
\frac{M_1}{t-M_1^2} \right|^2 \sim \frac{1}{M_1^2}
\end{equation}
for large $M_1$, where $M_1$ is the Bino mass.  The exact dependence
on $M_1$ is given in Fig.~\ref{fig:M1} for $\sqrt{s} = 500 \gev$ and
$m_{\selectron_R} = 200 \gev$. The dependence of $\sigma (e^+ e^-_R
\to \selectron^+_R \selectron^-_R)$ is also shown. We see that, in
stark contrast to the $\epem$ case, the $\emem$ cross section is large
and has a strong dependence on $M_1$ even for $M_1$ as high as ${\cal
O} (1\tev)$.  For example, for $M_1 = 700\gev$, the $1\sigma$
statistical error from a cross section measurement with $L=1\ifb$ is
$\Delta M_1 \approx 20 \gev$.  In addition, once $M_1$ is measured,
$M_2$ may be determined through the process $e^-_L e^-_L \to
\selectron^-_L \selectron^-_L$.  Note that such large gaugino masses,
which are possible in the Higgsino region in gravity-mediated models
and in other settings, are likely to be extremely difficult to measure
accurately by other means.  At the same time, these measurements are
extremely useful, for example, for testing gaugino mass unification.

\begin{figure}[t]
\centerline{\epsfig{file=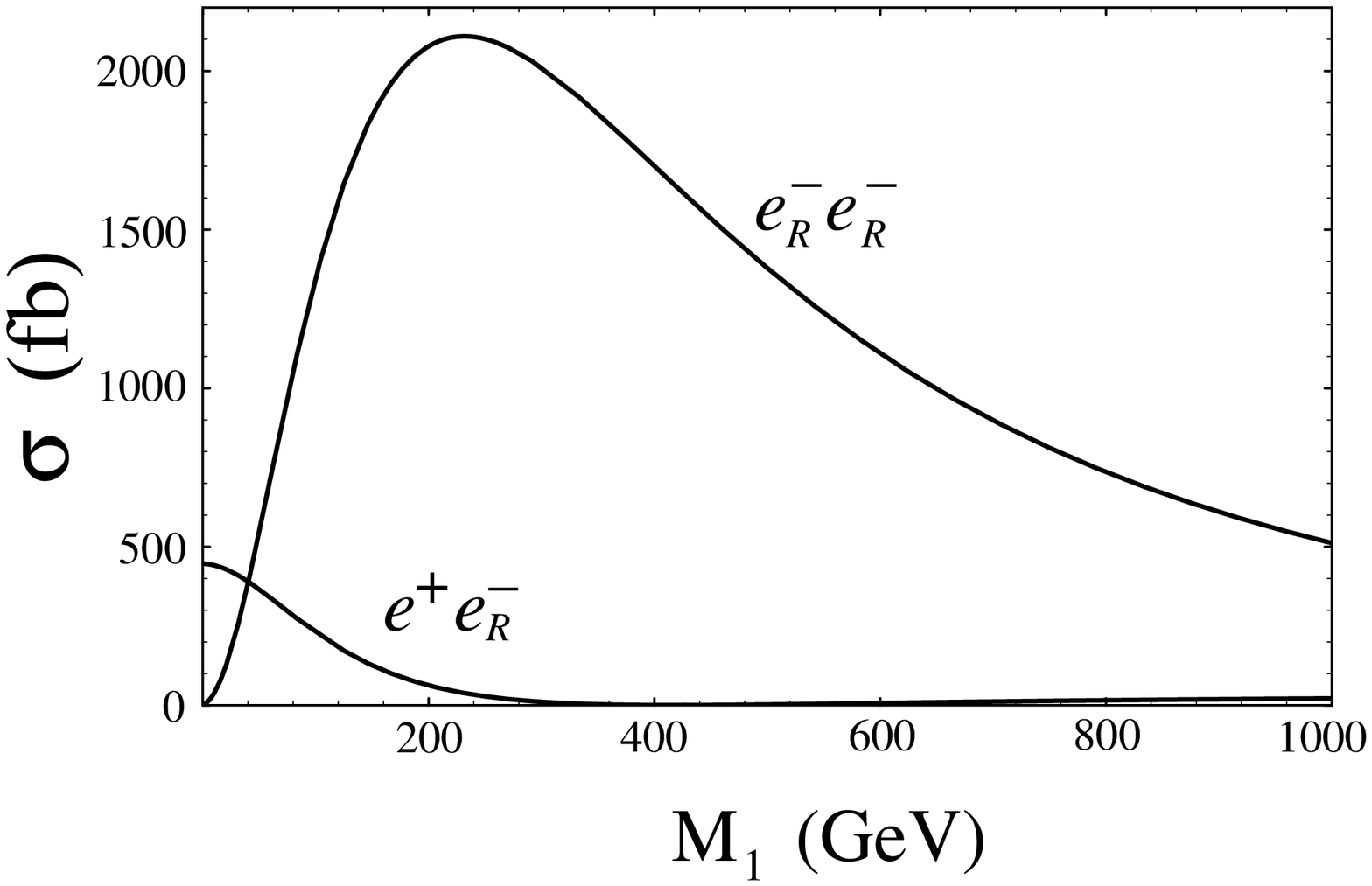,width=0.9\textwidth}}
\vspace*{0.1in}
\fcaption{Cross sections for $\sigma(e^-_R e^-_R \to \selectron^-_R
\selectron^-_R)$ and $\sigma(e^+e^-_R \to \selectron^+_R
\selectron^-_R)$ as functions of the Bino mass $M_1$ for
$m_{\selectron_R} = 200\gev$ and $\sqrt{s} = 500 \gev$. The
$t$-channel mass insertion for the $\emem$ case leads to large cross
sections, even for $M_1 \sim {\cal O} (1\tev)$.}
\label{fig:M1}
\end{figure}

\section{Conclusions}

It is clear that the possibilities for general studies of SUSY at an
$\epem$ collider cannot be matched by an $\emem$ collider.  However,
given that the $\emem$ mode is experimentally a relatively simple
extension of any linear collider program and is also motivated by the
desire for high energy $e\gamma$ and $\gamma\gamma$ studies, it is
certainly worth addressing what additional information the $\emem$
mode might bring to precision SUSY studies.

In this study, we have highlighted two possible applications.  First,
as a result of the fact that the scalar superpartners present in SUSY
theories have an associated handedness, $\emem$ colliders may enable
one to measure slepton masses through threshold scans with far greater
precision than in the $\epem$ mode.  Such high precision measurements
are useful for measuring $\tan\beta$, and, for example, may also allow
one to be sensitive to small radiative effects.\cite{CFP} 

It is also worth noting that such studies require far less luminosity
than the corresponding studies in the $\epem$ mode.  At present, most
studies of SUSY at linear colliders assume integrated luminosities of
$\gtrsim 20 \ifb$.  In addition, these studies often assume beam
energies and polarizations that are optimized for the particular study
at hand.  While it is clear that not all of these analyses may be
conducted simultaneously, systematic attempts to determine how best to
distribute the luminosity have not been undertaken, and, in any case,
may be premature, given the strong dependence on the actual
superpartner spectrum realized in nature.  However, in the event that
practical limitations on luminosity become relevant, novel studies
requiring only weeks of beam time may prove particularly attractive.

In addition, we have shown that the extraordinarily clean environment
of $\emem$ colliders leads to striking sensitivity in probes of
supersymmetric flavor structure through lepton flavor violation.  In
such studies, an accurate knowledge of beam polarization is crucial.
Finally, note that, for concreteness, we have concentrated on
scenarios with stable neutralinos as the lightest supersymmetric
particles.  However, in other scenarios, such as gauge-mediated
scenarios, the signals typically become much more spectacular, and the
results given above only improve.

\nonumsection{Acknowledgements}

I am grateful to the organizers, especially C.~Heusch and N.~Rogers,
for a stimulating and enjoyable conference and thank N. Arkani-Hamed,
H.--C. Cheng, L. Hall, and M. Peskin for conversations and
collaborations related to the work presented here.  This work was
supported in part by the Director, Office of Energy Research, Office
of High Energy and Nuclear Physics, Division of High Energy Physics of
the DOE under Contracts DE--AC03--76SF00098 and by the NSF under grant
PHY--95--14797.
 
\nonumsection{References}

\end{document}